\documentclass[twocolumn]{aastex7}
\usepackage{hyperref}
\usepackage{CJK}
\usepackage{graphicx}	
\usepackage{amsmath}	
\usepackage{amssymb}	
\usepackage{float}
\usepackage{graphicx}
\usepackage{subfigure}
\usepackage{multirow}

\usepackage{xcolor} 

\usepackage{hyperref}  

\submitjournal{ApJS}

\begin{document}
\begin{CJK*}{UTF8}{gbsn}

\title{Optimizing Long-term Variability of AGN Light Curves. I. A Case Study with ZTF Observations in the EGS Field}
\shortauthors{Lin et al.}

\correspondingauthor{Zhen-Ya Zheng,Bin Ma}
\email{zhengzy@shao.ac.cn,mabin3@mail.sysu.edu.cn}

\author[0000-0002-3134-9526]{Jiaqi Lin
({\CJKfamily{gbsn}林家琪})}
\affiliation{School of Physics and Astronomy, Sun Yat-sen University, Zhuhai 519082, People's Republic of China}
\affiliation{Shanghai Astronomical Observatory, Chinese Academy of Sciences, 80 Nandan Road, Shanghai, 200030, People's Republic of China}
\affiliation{CSST Science Center for the Guangdong-Hong Kong-Macau Greater Bay Area, Zhuhai 519082, People's Republic of China}
\email{linjiaqi@shao.ac.cn}

\author[0000-0002-9634-2923]{Zhen-Ya Zheng({\CJKfamily{gbsn}郑振亚})} 
\affiliation{Shanghai Astronomical Observatory, Chinese Academy of Sciences, 80 Nandan Road, Shanghai, 200030, People's Republic of China}
\email{zhengzy@shao.ac.cn}

\author[0000-0002-6077-6287]{Bin Ma({\CJKfamily{gbsn}马斌})} 
\affiliation{School of Physics and Astronomy, Sun Yat-sen University, Zhuhai 519082, People's Republic of China}
\affiliation{CSST Science Center for the Guangdong-Hong Kong-Macau Greater Bay Area, Zhuhai 519082, People's Republic of China}
\email{mabin3@mail.sysu.edu.cn}

\author[0000-0003-3787-0790]{Lin Long({\CJKfamily{gbsn}龙琳})}
\affiliation{Shanghai Astronomical Observatory, Chinese Academy of Sciences, 80 Nandan Road, Shanghai, 200030, People's Republic of China}
\affiliation{School of Astronomy and Space Sciences, University of Chinese Academy of Sciences, No. 19A Yuquan Road, Beijing 100049, People's Republic of China}
\email{longlin@shao.ac.cn}

\author[0000-0003-0962-2861]{Yangfan Xie
({\CJKfamily{gbsn}谢扬帆})}
\affil{Center For Astrophysics and Great Bay Center of National Astronomical Data Center, \\ Guangzhou University, Guangzhou, Guangdong, 510006, People's Republic of China}
\email{xyftokerbal@gmail.com}

\author[0009-0007-2850-9908]{Pu Lin
({\CJKfamily{gbsn}林浦})}
\affiliation{School of Physics and Astronomy, Sun Yat-sen University, Zhuhai 519082, People's Republic of China}
\affiliation{CSST Science Center for the Guangdong-Hong Kong-Macau Greater Bay Area, Zhuhai 519082, People's Republic of China}
\email{linp@mail2.sysu.edu.cn}

\author[0000-0003-3987-0858]{Ruqiu Lin
({\CJKfamily{gbsn}林如秋})}
\affiliation{Shanghai Astronomical Observatory, Chinese Academy of Sciences, 80 Nandan Road, Shanghai, 200030, People's Republic of China}
\affiliation{School of Astronomy and Space Sciences, University of Chinese Academy of Sciences, No. 19A Yuquan Road, Beijing 100049, People's Republic of China}
\email{linruqiu20@mails.ucas.ac.cn}

\author[0000-0002-0539-8244]{Xiang Ji({\CJKfamily{gbsn}吉祥})} 
\affiliation{Shanghai Astronomical Observatory, Chinese Academy of Sciences, 80 Nandan Road, Shanghai, 200030, People's Republic of China}
\email{jixiang@shao.ac.cn}



\begin{abstract}

Optical variability is a key observational probe for studying the accretion dynamics and central engine physics of Active Galactic Nuclei (AGNs). The quality and completeness of light curves have a direct impact on variability studies, particularly for faint AGNs and high-redshift AGNs. To improve the quality of long-term light curves for AGNs, we bin and stack multi-epoch images balancing the image depths and temporal resolution. As a case study, we apply this method to Zwicky Transient Facility (ZTF) observations in the Extended Groth Strip (EGS) field, where the overlapping region covers an area of about 370 arcmin$^2$ and includes $g$-band and $r$-band data taken from March 2018 to December 2024. The co-added images are approximately 2.0 to 2.5 magnitudes deeper than the ZTF single-epoch images. With co-added images, we construct light curves for 73 AGNs in the EGS field. Compared to the traditional ZTF light curves, our light curves maintain consistent long-term variability trends but with higher photometric precision. Furthermore, this method can help detect AGNs with weak variability which are missed from the traditional ZTF data due to the noisy light curves or below the detection limit in ZTF's single-epoch exposure. Among the 73 AGNs, the majority exhibit a bluer-when-brighter (BWB) trend on long-term timescales, which is consistent with previous studies. This work offers insights for optimizing AGN light curves in both current and upcoming all-sky time-domain surveys.
 
\end{abstract}




\section{Introduction} 
\label{1}

It is well known that Active Galactic Nuclei (AGNs) usually exhibit variability across the entire electromagnetic spectrum, with timescales ranging from hours to years \citep{Vanden-2004ApJ,Caplar-2017ApJ,Padovani-2017A&ARv}. In the optical band, quasars typically exhibit variability on timescales of several years, while showing relatively weak variations over shorter periods ranging from days to months \citep{Vanden-2004ApJ,MacLeod-2010ApJ,Stone-2022MNRAS,Li-2023ApJ}. Optical variability is a critical observational probe for studying the central physical structure and dynamics of AGNs \citep{Hook-1994MNRAS,Wilhite-2008MNRAS}.

Over the past decades, optical time-domain surveys have significantly advanced. Projects such as the Sloan Digital Sky Survey \citep[SDSS;][]{York-2000AJSDSS}, the Panoramic Survey Telescope and Rapid Response System \citep[Pan-STARRS;][]{Chambers-2016arXivpanstarr}, and the Zwicky Transient Facility \citep[ZTF;][]{Bellm-2019PASPZTF,Masci-2019PASPZTF} have enabled researchers to systematically investigate the connections between AGN variability and their physical parameters. Using extensive light-curve datasets, researchers have established statistical relationships between variability parameters (e.g., amplitude and timescales) and physical quantities such as redshift, rest-frame wavelength, luminosity, and black hole mass \citep{Vanden-2004ApJ,Wilhite-2008MNRAS,Kelly-2009ApJ,MacLeod-2010ApJ,Kozowski-2010ApJ,Zuo-2012ApJ,Simm-2016A&A,Caplar-2017ApJ,Stone-2022MNRAS}, which can provide critical constraints on the dynamical mechanisms of accretion processes. 
Variability studies have also provided crucial insights into extreme accretion processes and multi-scale interactions. For instance, some AGNs exhibit spectral type changes or the disappearance and reappearance of broad-line regions over timescales of months to years, typically accompanied by dramatic variability features. These objects are known as changing-look AGNs \citep{MacLeod-2016MNRAS,Yang-2018ApJ,Ricci-2023NatAs,Veronese-2024A&A,Guo-2024ApJS}, which are thought to be driven by mechanisms such as rapid changes in accretion rate, dynamic obscuration by dust structures, or transient accretion processes triggered by tidal disruption events. Additionally, quasi-periodic variability detected in a small fraction of AGNs may be associated with the orbital dynamics of supermassive black hole binary systems \citep{Graham-2015-Nature,Graham-MNRAS,Zheng-2016ApJ,Liu-2019ApJ}.

Although the variability of AGNs has been extensively studied, its physical origin remains unclear \citep{Simm-2016A&A}. Challenges include the limited sample size and the scarcity of high-quality light curves, particularly for AGNs that are observationally faint, such as low-luminosity AGNs and high-redshift AGNs. The former suffer from intrinsically low luminosity. The latter are dimmed by cosmological effects, including luminosity distance attenuation ($d_L^{-2}$) and bandpass shifting. Both of these populations are constrained by the detection thresholds of single exposures. To improve the quality of AGN light curves and increase sample size, it is essential to enhance image depth.

Image stacking is an effective method for enhancing image depth \citep{White-2007ApJ,Zibetti-2007ApJ,Zackay-2017ApJ-1,Zackay-2017ApJ-2}, as it improves the detection of faint sources by combining multi-epoch astronomical observations. For high-redshift AGNs, the observed variability timescales are stretched due to cosmological time dilation \citep{Hawkins-2010MNRAS,Dai-2012PhRvL}, which also benefits from image stacking. Through stacking multi-epoch SDSS Stripe 82 images, \citet{Abazajian-2009ApJS,Jiang-2009AJ,Huff-2014MNRAS,Jiang-2014ApJS,Annis-2014ApJ} generated deeper optical co-added images, enabling scientific studies of fainter magnitudes and higher redshifts. \citet{Jiang-2008AJ,Jiang-2009AJ,McGreer-2013ApJ} successfully identified high-redshift quasars at $z$ $>$ 5 with co-added SDSS Stripe 82 images. Furthermore, temporal stacking of adjacent exposures has been applied to mid-infrared data \citep{Meisner-2018AJ,Meisner-2023AJ}. By stacking the Wide-field Infrared Survey Explorer (WISE) and NEOWISE observations ($2010-2020$) at biannual intervals, \citet{Meisner-2023AJ} achieved a detection limit $\sim$1.3 magnitudes deeper than single exposures in W1 (3.4 $\mu$m) and W2 (4.6 $\mu$m) bands, enabling studies of fainter AGN with long-term mid-infrared variability.

Unlike WISE's uniform all-sky coverage with temporally regular sampling, the ground-based optical time-domain surveys such as the ZTF often exhibit irregular sampling. Here we develop a stacking methodology for ZTF's multi-epoch observations to enable comprehensive studies of AGN long-term optical variability while maintaining both sufficient image depth and temporal resolution. It should be noted that this increased depth comes at the cost of erasing short-term variability information, with the extent of this trade-off depending on the selected stacking-window width. As a case study, we select the Extended Groth Strip (EGS) field for our analysis due to its low extinction and abundance of deep multi-wavelength observations \citep{Davis-2007ApJ,Lotz-2008ApJ}. The EGS field provides deeper images and more complete catalogs, enabling the validation of our image-stacking-based optimization method for investigating AGN long-term variability.

The paper is organized as follows. We describe the ZTF data in the EGS field and our data processing procedure in Section \ref{2}. To balance the image depths and temporal resolution, we perform binning and stacking of ZTF single-epoch images, followed by photometry and light curve extraction on the co-added images. We then present the comparative analysis of the optimized light curves and the original ZTF light curves in Section \ref{3}. Finally, we provide discussion and summary in Section \ref{4} and Section \ref{5}, respectively. Throughout this study, we adopt $H_{0}$=70~$\rm{km\ s^{-1}\ Mpc^{-1}}$, $\rm{\Omega}_{m}=0.3$ and $\rm{\Omega_{\Lambda}=0.7}$.

\section{Data and Methods} 
\label{2}

\subsection{Data Description}
\label{2.1}

The EGS field is a deep-sky field observed by the Hubble Space Telescope (HST) Advanced Camera for Surveys (ACS) under GO Program 10134 (PI: M.Davis) \citep{Davis-2007ApJ}. Centered at $\alpha = 214.825^\circ, \delta = 52.823^\circ$, it covers a rectangular region of approximately $10.1 \, \text{arcmin} \times 70.5 \, \text{arcmin}$.

The ZTF, an optical time-domain survey, uses Palomar Observatory's 48-inch Schmidt telescope and a 576-megapixel camera with a 47 deg$^2$ field of view \citep{Bellm-2019PASPZTF,Masci-2019PASPZTF}. Since its observations began in 2018, ZTF has covered over 25,000 deg$^2$ of the sky, with some regions observed as many as 1,000 to 2,000 times in $g$-band and $r$-band, respectively. These multi-epoch observations facilitate image stacking to enhance image depth and enable detailed AGN variability studies.

\begin{figure*}[!htbp]
\centering
\includegraphics[width=1\linewidth]{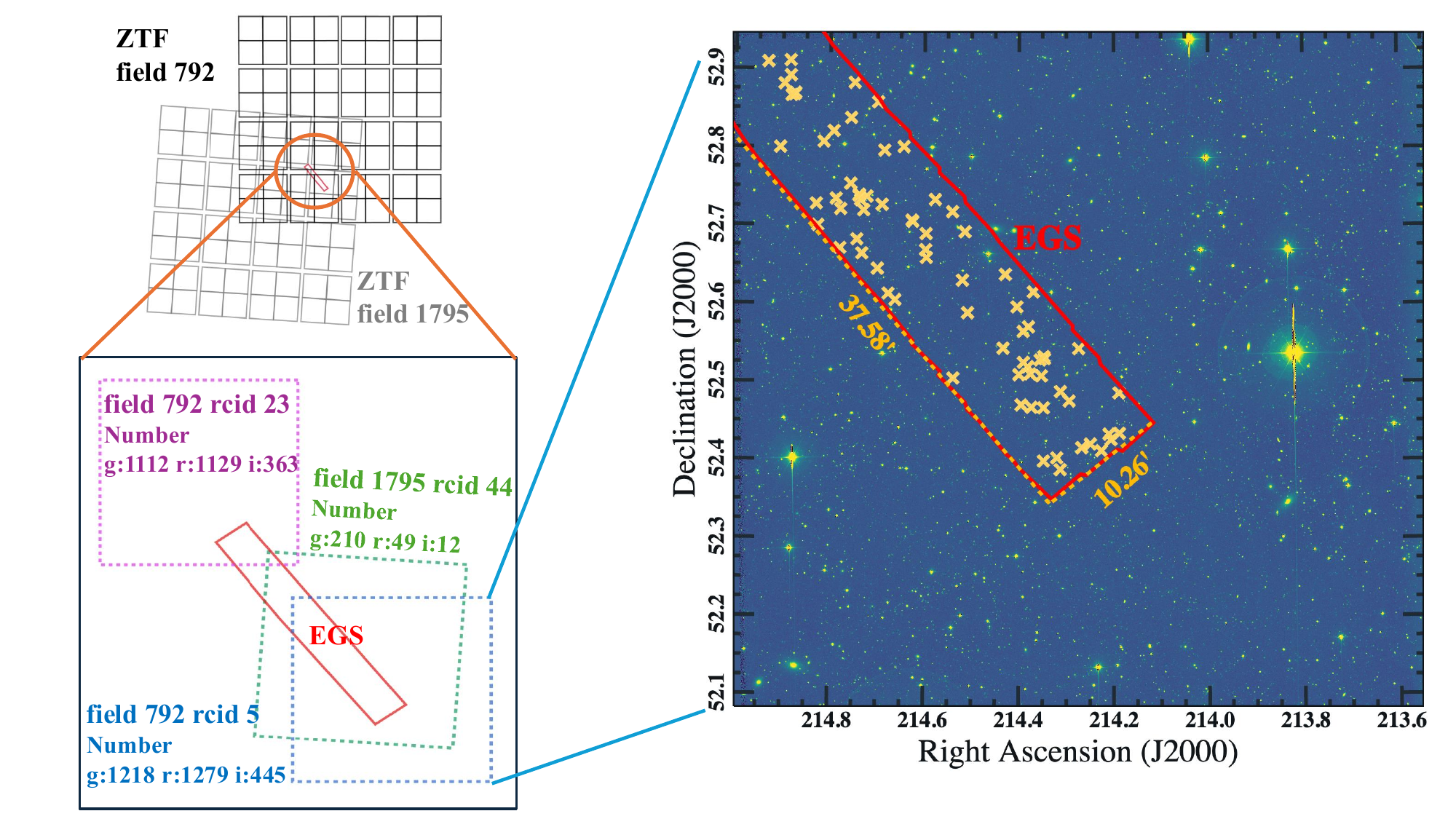}
\caption{Schematic diagram of the overlapping region between ZTF and the EGS field. The left panel shows the coverage of ZTF's three CCD-quadrants with the EGS field, along with the number of observations in the $g$, $r$, and $i$ bands from March 2018 to December 2024. The right panel illustrates the overlap between the blue CCD-quadrant (field 792 rcid 5) and the EGS field, with an overlapping area (marked by a solid red line) of approximately $10.2 \, \text{arcmin} \times 37.5 \, \text{arcmin}$. Yellow "x" symbols represent the 73 AGNs extracted in this study. Axes are in degrees.}
\label{fig:ztf_region}
\end{figure*}

As shown in Figure \ref{fig:ztf_region}, the ZTF has two fields overlapping with the EGS field, namely field 792 and field 1795. Each field is imaged with 16 charge coupled devices (CCDs), each of which has 4 readout channels, resulting in a total of 64 CCD-quadrants per field. The CCD-quadrant is the basic image unit for the ZTF data processing pipeline, from which all scientific data products are derived. There are three CCD-quadrants overlapping with the EGS field, labeled as field 792 rcid 23 (purple), field 1795 rcid 44 (green), and field 792 rcid 5 (blue). We present the numbers of exposures for these three CCD-quadrants in the $g$, $r$, and $i$ bands from March 2018 to December 2024 in Figure \ref{fig:ztf_region}. Due to a small number of exposures, the CCD-quadrant of field 1795 rcid 44 is excluded. Similarly, the CCD-quadrant of field 792 rcid 23 is excluded due to its small overlap area with the EGS field. Ultimately, the CCD-quadrant of field 792 rcid 5 is selected for analysis, with an overlap area of approximately $10.2 \, \text{arcmin} \times 37.5 \, \text{arcmin}$ with the EGS field. This CCD-quadrant has accumulated 1218, 1279, and 445 exposures in the $g$, $r$, and $i$ bands, respectively.

To ensure the consistent quality of single-epoch images, we select science images with image-level quality flags (INFOBITS) set to 0, indicating no identified quality issues. Each science image has an exposure time of 30 s. After filtering, we obtain 1004 $g$-band, 1099 $r$-band and 201 $i$-band science images. Due to the limited $i$-band sample size, we exclude it from further analysis. Each science image is approximately $52 \, \text{arcmin} \times 52 \, \text{arcmin}$, and $3072 \, \times 3080 \,$ pixels with a pixel scale of 1.012 arcsec pixel$^{-1}$. The median full widths at half-maximum (FWHMs) are 2.20 arcsec ($g$-band) and 1.88 arcsec ($r$-band). The median 5 $\sigma$ limiting magnitudes are 20.58 mag ($g$-band) and 20.61 mag ($r$-band) in the AB system.


The FITS header of each ZTF science image contains key observational parameters, including seeing FWHM, limiting magnitude, and photometric zero-point. In addition, ZTF provides a corresponding mask image for each science image, which records pixel-level data quality information and identifies pixels affected by conditions such as saturation, cosmic rays, or bad pixels. This information is utilized in subsequent data processing workflows.

\subsection{Data Processing}
\label{2.2}

\begin{figure*}[!htbp]
\centering
\includegraphics[width=2 \columnwidth]{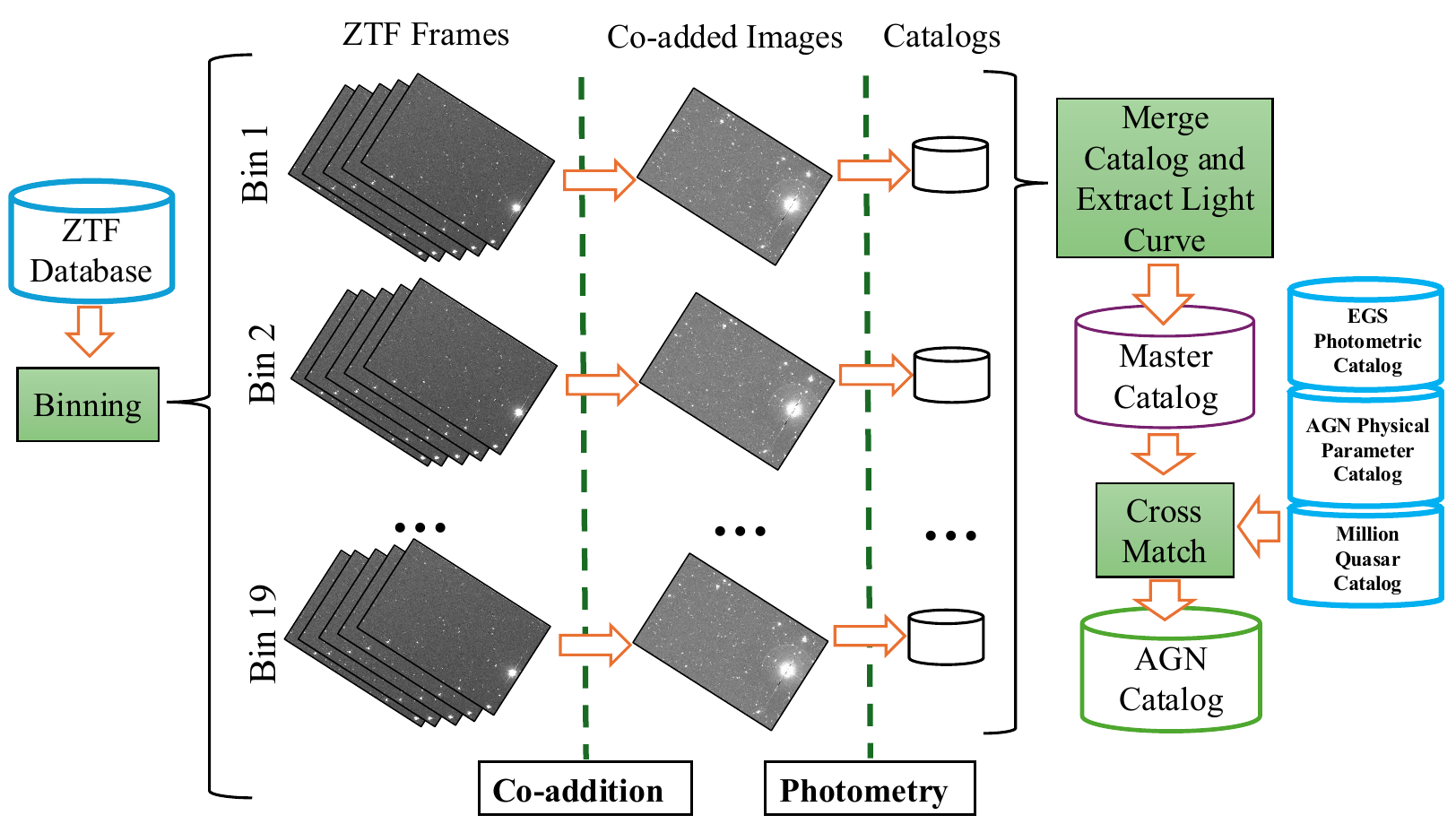}
\caption{Schematic diagram of the data processing workflow.}
\label{fig:flowchart}
\end{figure*}

To improve the quality of light curves for faint AGNs, we perform binning and stacking of ZTF single-epoch images balancing the image depths and temporal resolution. We then conduct photometry and light curve extraction on the co-added images. The data processing workflow of this study is illustrated in Figure \ref{fig:flowchart}.


\subsubsection{Data Binning}
\label{2.2.1}



To balance the image depths and temporal resolution, we bin the science images by observational cadence and stack the frames in each bin to generate a co-added image. Each co-added image represents the average of all frames in that bin, with its observation time defined as the median time of the constituent frames.

\begin{figure*}[!htbp]
\centering
\includegraphics[width=2 \columnwidth]{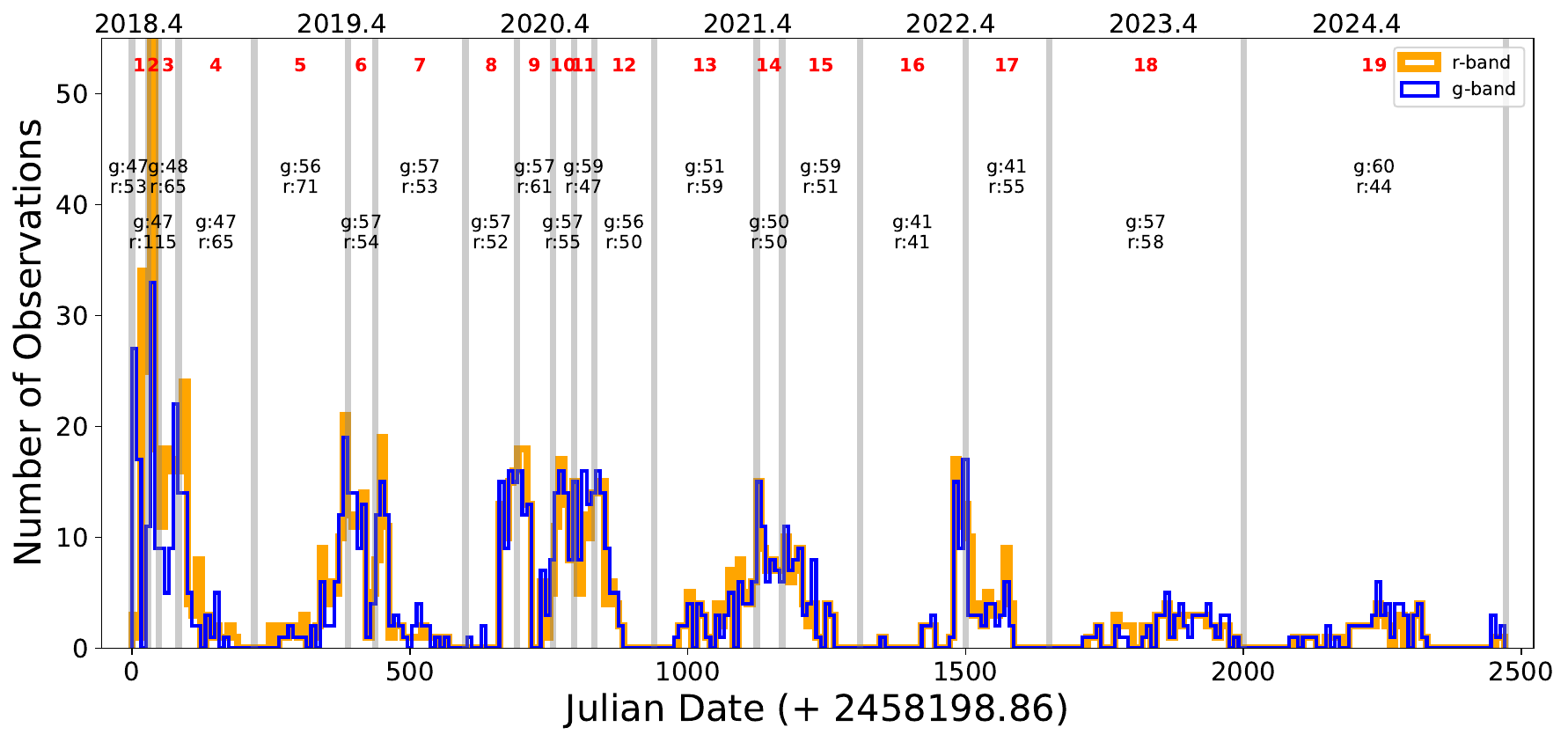}
\caption{Observation cadence of ZTF $g$-band (blue) and $r$-band (orange) exposures in field 792 rcid 5 from March 2018 to December 2024. The gray vertical lines represent the range of each bin, the red numbers indicate the labels of each bin, and the black numbers represent the number of exposures in the g-bang and $r$-band for each bin.}
\label{fig:bin}
\end{figure*}

As shown in Figure \ref{fig:bin}, the $g$-band and $r$-band observation cadences are generally synchronized. To align the cadence, we treat local peak intervals in observation counts as integrated units. To ensure sufficient S/N for detecting faint sources, we set a minimum threshold of 40 $g$-band or $r$-band images per bin. Typically, local peak intervals are divided into 2 to 3 bins, but the final two intervals (covering 2023 and 2024 data) have fewer exposures and are each assigned to a single bin. Notably, both $g$-band and $r$-band share identical time ranges for each bin, ensuring consistent temporal alignment in the co-added images. In total, we create 19 bins. The detailed information of each bin is listed in Table \ref{tab:1}.

\subsubsection{Co-addition Process}
\label{2.2.2}

The weighted averaging method optimizes image combination by assigning weights based on key observational conditions, including seeing, sky brightness, and atmospheric transparency \citep{Annis-2014ApJ,Jiang-2014ApJS}. This method effectively suppresses noise and enhances the S/N for faint objects. In this study, we use \emph{SWarp} \citep{Bertin-2010ascl}\footnote{http://www.astromatic.net/software/swarp} for image stacking and apply the weight proposed by \citet{Annis-2014ApJ}. The weight is defined as follows:

\begin{equation}{} 
W_{i(x,y)}  = \frac{T_{i}^{2} }{\textrm{FWHM}_{i}^{2}\sigma_{i(x,y)}^{2}},
\label{equ-1}
\end{equation} 
where $\textrm{FWHM}_{i}$ represents the seeing of the $i$-th image, $\sigma_{i(x,y)}$ denotes the sky noise per pixel, and T stands for the transparency. Additionally, we assign a weight of 0 to other anomalous pixels, such as bad pixels, those contaminated by cosmic rays, and those affected by satellite trails.

Before image stacking, it is necessary to perform sky background subtraction for each science image. We use the Python \emph{SEP}\citep{Barbary2016}\footnote{https://github.com/sep-developers/sep} package for sky background modeling. Specifically, we extract non-background pixel labels from the mask images provided by ZTF and apply them as the mask parameter in \emph{sep.Background}. The sky background in ZTF images is modeled using 256-pixel sampling boxes and 3-pixel smoothing filters. Finally, the modeled sky background is subtracted from the original science image.

Variations in atmospheric transparency and instrumental response can lead to inconsistent photometric throughput across science images. To ensure photometric consistency, we scale all science images to a fixed target zero-point, with each band aligned to a specific reference value. The target zero-point $ZP_{target}$ is typically determined by the median photometric zero-point of all available images \citep{Masci-2019PASPZTF}. We then calculate the flux scaling factor ($FLXSCALE = 10^{-0.4(ZP_{image} - ZP_{target} )}$) for each science image and record this parameter in the corresponding FITS header. During the image stacking process, \emph{SWarp} automatically retrieves the \emph{FLXSCALE} value from each image and applies the appropriate scaling factor to calibrate the relative flux before combining the images.

\begin{figure*}[!htbp]
\centering
\includegraphics[width=2 \columnwidth]{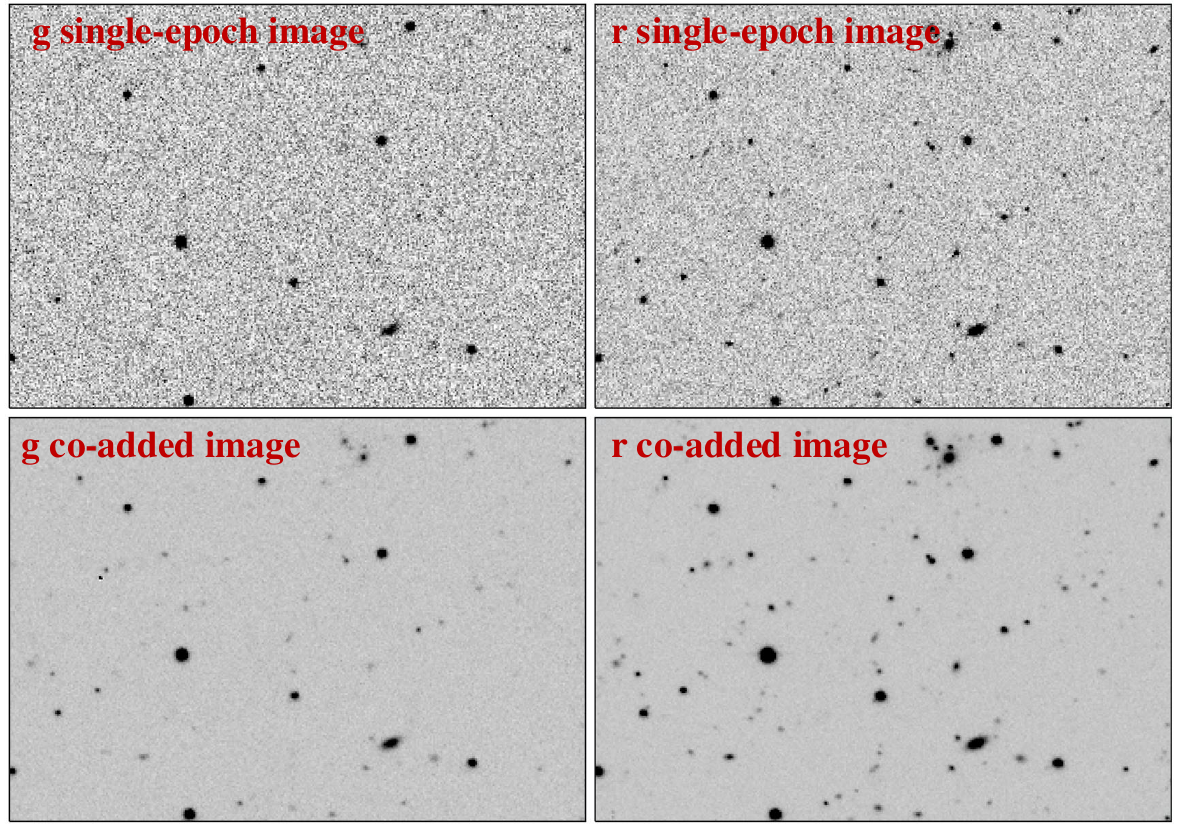}
\caption{A small 5.3 arcmin × 3.7 arcmin cutout centered at $\alpha = 214.35^{\circ } ,\delta = 52.56^{\circ }$ compares single-epoch (top) and co-added (bottom) images in two bands. All images are displayed using the same scale, contrast, and stretch.}
\label{fig:compare}
\end{figure*}

Finally, we use \emph{SWarp} to resample and combine images. In the \emph{SWarp} configuration, we set the resampling type (\emph{RESAMPLING\_TYPE}) to \emph{LANCZOS3}, which can effectively preserve image details and reduce aliasing effects \citep{Bertin-2010ascl}. The combination type (\emph{COMBINE\_TYPE}) is set to weighted averaging (\emph{WEIGHTED}), which is particularly suitable for detecting and measuring faint sources in well-weighted and homogeneous images. The output pixel scale is maintained at the native resolution of 1.012 arcsec pixel$^{-1}$. Following this processing pipeline, we produce 19 co-added images for each band.

\begin{figure*}[!htbp]
\centering
\includegraphics[width=\textwidth]{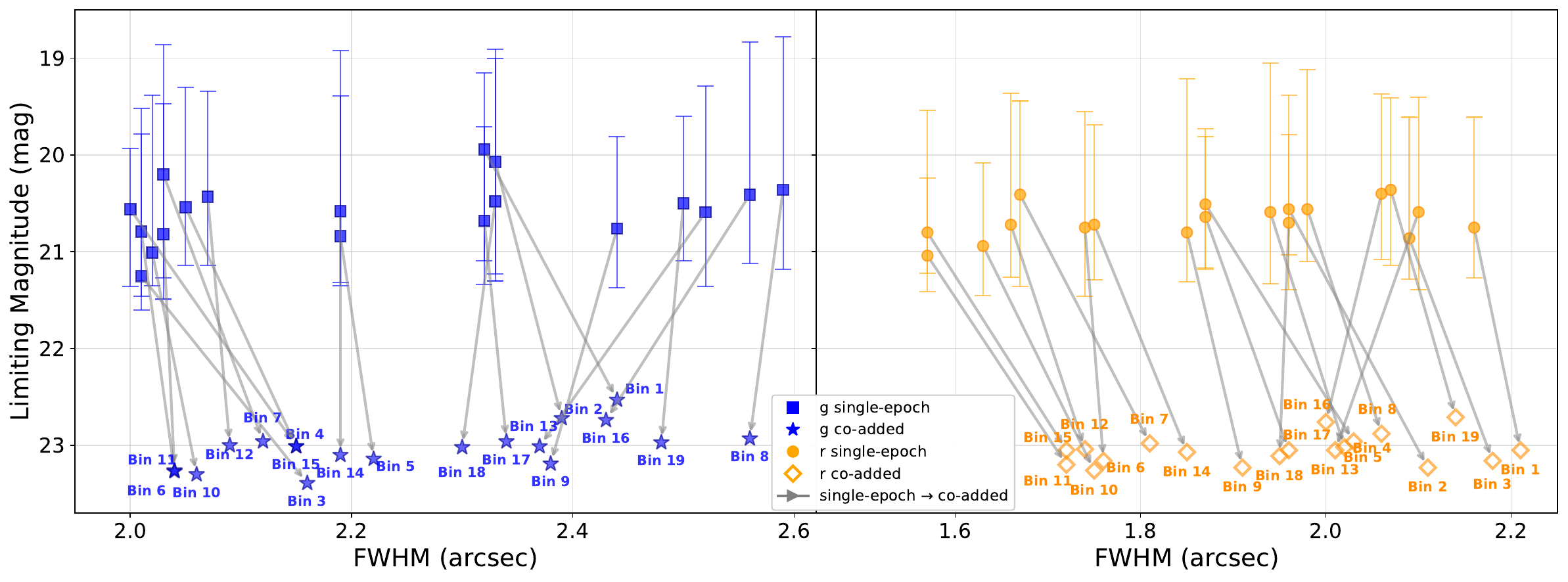}
\caption{The FWHM versus limiting magnitude for $g$-band (blue, left panel) and $r$-band (orange, right panel). Vertical bars represent the limiting magnitude ranges for single-epoch data. In the $g$-band, single-epoch medians are marked by squares and co-added values by stars. In the $r$-band, single-epoch medians are marked by circles and co-added values by diamonds. Gray arrows connect single-epoch to co-added median points. Each data point is labeled with its corresponding sample bin (Bin 1--19).}
\label{fig:fwhm_vs_lm}
\end{figure*}

As shown in Figure \ref{fig:compare}, the co-added images have lower noise and higher S/N compared to the single-epoch images, enhancing the detection of faint sources. We then compare the FWHM and limiting magnitude between single-epoch and co-added images for each bin in the $g$ and $r$ bands, as presented in Table \ref{tab:1} and Figure \ref{fig:fwhm_vs_lm}. The FWHM is measured from unsaturated, isolated, and bright point sources (with magnitudes between 14 and 18, S/N$>$20). The results show that the FWHM of co-added images is close to the median FWHM of single-epoch images in each bin, with the $g$-band FWHM slightly larger than that of the $r$-band. The limiting magnitude is derived from Point Spread Function (PSF) photometry, based on the median magnitude of sources with S/N values between 4.5 and 5.5 to estimate an approximate 5$\sigma$ magnitude limit. The limiting magnitude improves significantly in co-added images, increasing by approximately 2.0 to 2.5 magnitudes (e.g., from 19.94 to 22.53 in g-band Bin 1). This aligns with the theoretical $\sqrt{N}$ scaling of the S/N under Poisson noise, predicting a depth increase of approximately $2.5 \log_{10}(\sqrt{40}) \approx 2.0$ magnitudes when co-adding $N=40$ exposures.

\begin{table*}
\begin{center}
\renewcommand{\arraystretch}{1.2}
\setlength{\tabcolsep}{2.2pt}
\caption{Statistical values for each bin. The Julian Date represents the median observation time within the bin. For single-epoch images, the FWHM and limiting magnitude are reported as median values (with min, max in parentheses) within each bin. For co-added images, the median values are provided.}
\label{tab:1}

\begin{tabular*}{\linewidth}{@{\extracolsep{\fill}}cccccccc@{}}
\hline
\multirow{3}{*}{Bin}    & \multirow{3}{*}{Julian Date} & \multirow{3}{*}{Band} & \multirow{3}{*}{Stacking Number} & \multicolumn{2}{c}{FWHM (arcsec)}                                           & \multicolumn{2}{c}{Limiting Magnitude (mag)}                                \\
                        &                              &                       &                                  & \multicolumn{1}{l}{Single-epoch Image} & \multicolumn{1}{l}{Co-added Image} & \multicolumn{1}{l}{Single-epoch Image} & \multicolumn{1}{l}{Co-added Image} \\
                        &                              &                       &                                  & Median (Min,Max)                        &                              & Median (Min,Max)                        &                              \\ \hline
\multirow{2}{*}{Bin 1}  & \multirow{2}{*}{2458213.36}  & $g$                   & 47                               & 2.32 (1.66, 4.16)                      & 2.44                               & 19.94 (19.15, 21.34)                   & 22.53                              \\
                        &                              & $r$                   & 53                               & 2.16 (1.62, 3.08)                      & 2.21                               & 20.75 (19.61, 21.27)                   & 23.05                              \\
\multirow{2}{*}{Bin 2}  & \multirow{2}{*}{2458234.81}  & $g$                   & 47                               & 2.33 (1.75, 3.77)                      & 2.39                               & 20.07 (18.91, 21.23)                   & 22.72                              \\
                        &                              & $r$                   & 115                              & 1.96 (1.39, 3.50)                      & 2.11                               & 20.56 (19.79, 21.03)                   & 23.23                              \\
\multirow{2}{*}{Bin 3}  & \multirow{2}{*}{2458269.76}  & $g$                   & 48                               & 2.01 (1.70, 2.71)                      & 2.16                               & 21.25 (19.78, 21.60)                   & 23.39                              \\
                        &                              & $r$                   & 65                               & 2.09 (1.56, 4.82)                      & 2.18                               & 20.86 (19.61, 21.28)                   & 23.16                              \\
\multirow{2}{*}{Bin 4}  & \multirow{2}{*}{2458295.24}  & $g$                   & 47                               & 2.00 (1.73, 2.85)                      & 2.15                               & 20.56 (19.93, 21.36)                   & 23.01                              \\
                        &                              & $r$                   & 65                               & 1.87 (1.51, 3.68)                      & 2.03                               & 20.51 (19.81, 21.18)                   & 22.96                              \\
\multirow{2}{*}{Bin 5}  & \multirow{2}{*}{2458567.86}  & $g$                   & 56                               & 2.19 (1.67, 4.11)                      & 2.22                               & 20.84 (18.92, 21.35)                   & 23.14                              \\
                        &                              & $r$                   & 71                               & 1.94 (1.39, 4.82)                      & 2.02                               & 20.59 (19.05, 21.33)                   & 23.02                              \\
\multirow{2}{*}{Bin 6}  & \multirow{2}{*}{2458606.87}  & $g$                   & 57                               & 2.03 (1.65, 4.84)                      & 2.04                               & 20.82 (18.86, 21.49)                   & 23.26                              \\
                        &                              & $r$                   & 54                               & 1.74 (1.29, 2.56)                      & 1.76                               & 20.75 (19.55, 21.46)                   & 23.16                              \\
\multirow{2}{*}{Bin 7}  & \multirow{2}{*}{2458653.31}  & $g$                   & 57                               & 2.03 (1.56, 3.83)                      & 2.12                               & 20.20 (19.47, 21.27)                   & 22.96                              \\
                        &                              & $r$                   & 53                               & 1.67 (1.34, 3.98)                      & 1.81                               & 20.41 (19.44, 21.36)                   & 22.98                              \\
\multirow{2}{*}{Bin 8}  & \multirow{2}{*}{2458877.51}  & $g$                   & 57                               & 2.59 (1.90, 4.57)                      & 2.56                               & 20.36 (18.78, 21.18)                   & 22.93                              \\
                        &                              & $r$                   & 52                               & 1.98 (1.47, 4.66)                      & 2.06                               & 20.56 (19.12, 21.10)                   & 22.88                              \\
\multirow{2}{*}{Bin 9}  & \multirow{2}{*}{2458907.02}  & $g$                   & 57                               & 2.44 (1.91, 3.77)                      & 2.38                               & 20.76 (19.81, 21.37)                   & 23.19                              \\
                        &                              & $r$                   & 61                               & 1.85 (1.41, 4.08)                      & 1.91                               & 20.80 (19.21, 21.31)                   & 23.23                              \\
\multirow{2}{*}{Bin 10} & \multirow{2}{*}{2458974.25}  & $g$                   & 57                               & 2.02 (1.72, 3.03)                      & 2.06                               & 21.01 (19.38, 21.35)                   & 23.30                              \\
                        &                              & $r$                   & 55                               & 1.63 (1.27, 2.91)                      & 1.75                               & 20.94 (20.08, 21.45)                   & 23.26                              \\
\multirow{2}{*}{Bin 11} & \multirow{2}{*}{2459012.82}  & $g$                   & 59                               & 2.01 (1.69, 3.97)                      & 2.04                               & 20.79 (19.52, 21.46)                   & 23.27                              \\
                        &                              & $r$                   & 47                               & 1.57 (1.30, 3.60)                      & 1.72                               & 21.04 (20.24, 21.41)                   & 23.20                              \\
\multirow{2}{*}{Bin 12} & \multirow{2}{*}{2459045}     & $g$                   & 56                               & 2.07 (1.64, 3.25)                      & 2.09                               & 20.43 (19.34, 21.14)                   & 23.00                              \\
                        &                              & $r$                   & 50                               & 1.66 (1.38, 2.36)                      & 1.74                               & 20.72 (19.36, 21.26)                   & 23.04                              \\
\multirow{2}{*}{Bin 13} & \multirow{2}{*}{2459275.85}  & $g$                   & 51                               & 2.52 (1.62, 4.57)                      & 2.37                               & 20.59 (19.29, 21.36)                   & 23.01                              \\
                        &                              & $r$                   & 59                               & 2.10 (1.35, 4.13)                      & 2.01                               & 20.59 (19.40, 21.39)                   & 23.05                              \\
\multirow{2}{*}{Bin 14} & \multirow{2}{*}{2459341.12}  & $g$                   & 50                               & 2.19 (1.75, 4.31)                      & 2.19                               & 20.58 (19.39, 21.32)                   & 23.10                              \\
                        &                              & $r$                   & 50                               & 1.75 (1.45, 3.37)                      & 1.85                               & 20.72 (19.69, 21.29)                   & 23.07                              \\
\multirow{2}{*}{Bin 15} & \multirow{2}{*}{2459399.24}  & $g$                   & 59                               & 2.05 (1.67, 2.83)                      & 2.15                               & 20.54 (19.30, 21.14)                   & 23.01                              \\
                        &                              & $r$                   & 51                               & 1.57 (1.30, 2.41)                      & 1.72                               & 20.80 (19.54, 21.22)                   & 23.05                              \\
\multirow{2}{*}{Bin 16} & \multirow{2}{*}{2459684.29}  & $g$                   & 41                               & 2.56 (1.75, 3.93)                      & 2.43                               & 20.41 (18.83, 21.12)                   & 22.74                              \\
                        &                              & $r$                   & 41                               & 2.06 (1.37, 3.56)                      & 2.00                               & 20.40 (19.37, 21.08)                   & 22.76                              \\
\multirow{2}{*}{Bin 17} & \multirow{2}{*}{2459735.77}  & $g$                   & 41                               & 2.32 (1.68, 3.34)                      & 2.34                               & 20.68 (19.71, 21.09)                   & 22.96                              \\
                        &                              & $r$                   & 55                               & 1.87 (1.38, 3.50)                      & 1.96                               & 20.64 (19.73, 21.17)                   & 23.05                              \\
\multirow{2}{*}{Bin 18} & \multirow{2}{*}{2460082.33}  & $g$                   & 57                               & 2.33 (1.82, 4.32)                      & 2.30                               & 20.48 (19.00, 21.30)                   & 23.02                              \\
                        &                              & $r$                   & 58                               & 1.96 (1.43, 3.85)                      & 1.95                               & 20.70 (19.38, 21.39)                   & 23.11                              \\
\multirow{2}{*}{Bin 19} & \multirow{2}{*}{2460448.28}  & $g$                   & 60                               & 2.50 (1.85, 4.09)                      & 2.48                               & 20.50 (19.60, 21.09)                   & 22.97                              \\
                        &                              & $r$                   & 44                               & 2.07 (1.59, 3.04)                      & 2.14                               & 20.36 (19.41, 21.14)                   & 22.71                              \\ \hline
\end{tabular*}
\end{center}
\end{table*}

\subsubsection{Light Curve Extraction}
\label{2.2.3}

To extract the light curves for all sources on the co-added images, we first perform photometry, then calibrate the fluxes, and finally merge all catalogs.

We use \emph{SExtractor}\footnote{https://www.astromatic.net/software/sextractor} \citep{Bertin-1996A&AS-sex} and \emph{PSFEx}\footnote{https://www.astromatic.net/software/psfex} \citep{Bertin-2011ASPC-PSFEx} to perform aperture photometry and PSF photometry on each co-added image. We adopt a detection threshold of 1.5 $\sigma$ per pixel (\emph{DETECT\_THRESH}) and a minimum area of 5 connected pixels (\emph{DETECT\_MINAREA}) to ensure source detection completeness and reliability. Apertures of 2, 3, 4 and 6 pixels in radius are used for aperture photometry. To balance measurements between bright and faint sources, we select the 3-pixel radius aperture magnitude as the default magnitude. Source positions are measured via \emph{XWIN\_WORLD} and \emph{YWIN\_WORLD} centroids. PSF photometry is more optimal for measuring faint sources \citep{Becker-2007PASP}. Additionally, the light curves provided by ZTF are constructed based on PSF photometry \citep{Masci-2019PASPZTF}. Therefore, in our subsequent light curve extraction, we prioritize PSF photometry results. Only when a measurement point in the entire light curve cannot be fitted using PSF photometry do we switch to the default aperture photometry result. For \emph{PSFEx}, we set \emph{BASIS\_TYPE} to \emph{PIXEL\_AUTO} and define the \emph{PSF\_SIZE} as 25×25 pixels. The PSF variation is modeled using a second-degree polynomial by setting \emph{PSFVAR\_DEGREES} to 2.

In flux calibration process, we use the Pan-STARRS1 DR1 (PS1) MeanObject catalog \citep{Chambers-2016arXivpanstarr,Flewelling-2020ApJS} as the reference catalog. The calibration procedure involves cross-matching the instrumental catalog with the reference catalog using a 1-arcsec matching radius and selecting sources with magnitudes between 14 and 18. After excluding 3$\sigma$ outliers, we use the least-squares method to calculate the magnitude zero-point.

We employ a two-stage strategy for merging catalogs. We first merge the catalogs from all bins within each band to create an initial master catalog. For source matching, we utilize the \emph{search\_around\_sky}\footnote{https://docs.astropy.org/en/stable/api/\\astropy.coordinates.SkyCoord.html\\\#astropy.coordinates.SkyCoord.search\_around\_sky} function from Astropy's SkyCoord module, applying a matching radius of 1 arcsec. Since sources may be detected multiple times with slightly varying positions, we compute the average position of these detections to determine the final coordinates. After merging the catalogs for each band independently, we repeat this process to produce a final master catalog that integrates data from both bands. Using this master catalog, we index the source positions to extract their light curve data.

\subsection{AGN Catalog in the EGS Field}
\label{2.3}

\begin{figure*}[!htbp]
\centering
\includegraphics[width=2 \columnwidth]{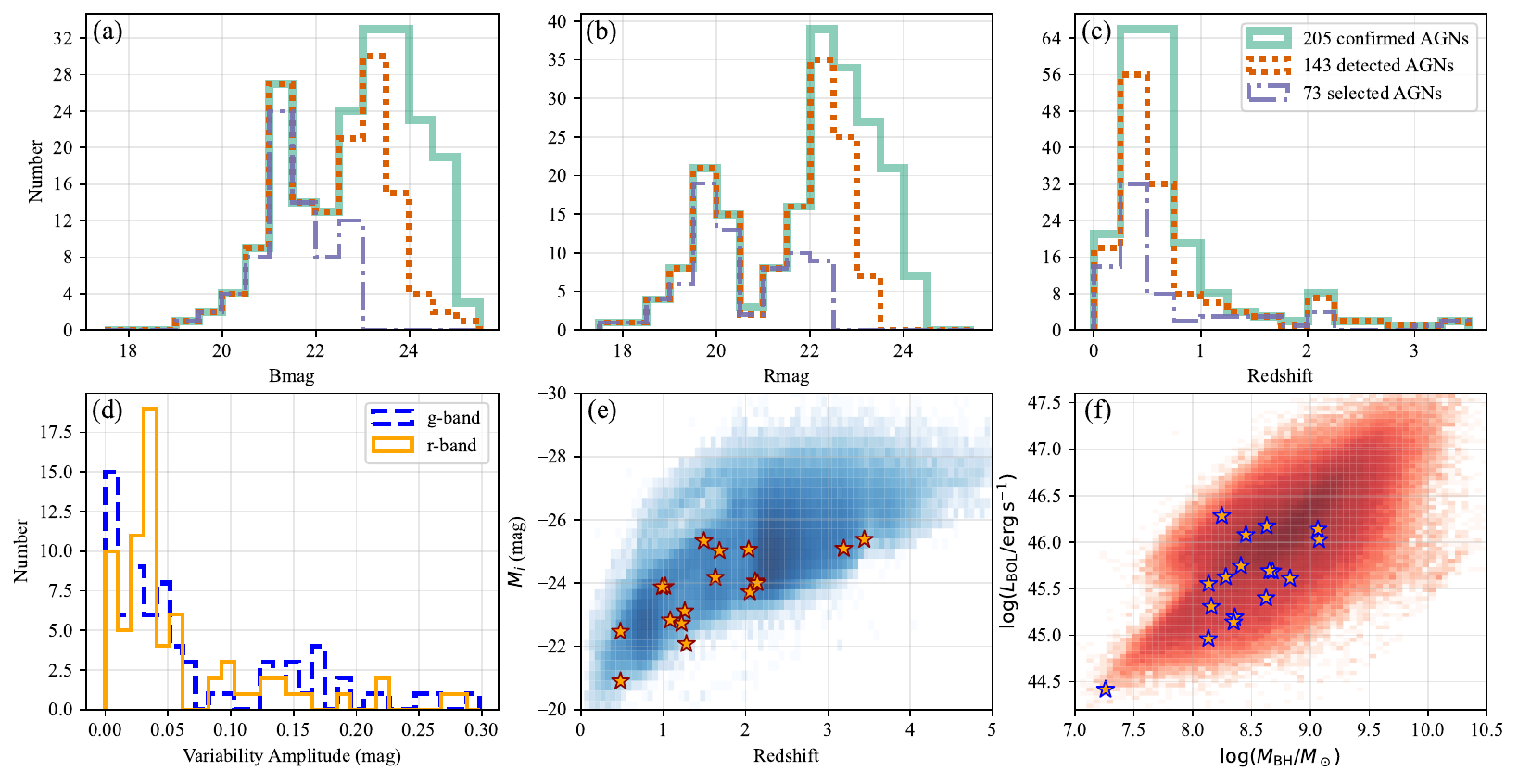}
\caption{Panels (a)--(c): Magnitude (from MQC) and redshift distributions for the 205 confirmed AGNs, 143 detected AGNs, and the final sample of 73 selected AGNs in the EGS field. Here, Bmag and Rmag denote the optical blue and red bands, respectively. The specific filters and central wavelengths vary by observational source. Panel (d): Variability amplitude for the final catalog of 73 AGNs. Panels (e)--(f): For a subset of 17 AGNs (marked with stars), the distributions of (e) redshift versus absolute magnitude and (f) black hole mass versus bolometric luminosity are shown. In both panels (e) and (f), the blue and red points represent the two-dimensional parameter distribution from the larger AGN Physical Parameter Catalog (approximately 280,000 quasars) presented in \citet{Szymon-2017ApJS}.}
\label{fig:AGNcat_dis}
\end{figure*}

To validate the advantages of our method optimized for studying long-term variability of AGN light curves, we first need to obtain a deep AGN catalog for the EGS field (as outlined with the red rectangle on the right panel of Figure \ref{fig:ztf_region}, covering an area of about 370 arcmin$^2$). We start with the the Million Quasar Catalog \citep[MQC; version 8.0, released on June 30, 2023;][]{Flesch-2023OJAp}, which contains a total of 1,021,800 AGNs with information such as their positions, types, redshifts, red and blue optical magnitudes (denoted as Rmag and Bmag, respectively). The precise filters and wavelengths for the red and blue bands depend on the source, as specified by letter codes in the ``Comment'' field. For example, a ``Comment'' value of ``g'' indicates that the blue magnitude corresponds to the SDSS-type green filter (4900\,\AA) and the red magnitude corresponds to the SDSS $r$ filter (6200\,\AA). There are 205 confirmed AGNs from the MQC in the EGS field. 

The magnitude and redshift distributions of the 205 AGNs are presented in panels (a)--(c) of Figure \ref{fig:AGNcat_dis}. The Bmag range from 19.3 to 25.13, the Rmag range from 17.61 to 24.07, and the redshifts range from 0.161 to 3.462. The histograms of both Bmag and Rmag exhibit a bimodal distribution. The MQC incorporates data from multiple survey sources (e.g., APM, SDSS, and Pan-STARRS), which were obtained under non-uniform observational conditions. These datasets are integrated from surveys with varying detection limits, classification criteria, and photometric corrections \citep{{Flesch-2023OJAp}}. Consequently, the observed bimodal distribution is likely attributable to this heterogeneity in the data sources.

Compared to our co-added ZTF data, the MQC of 205 AGNs exhibits greater depth and completeness. Among these 205 AGNs, we detect 143 AGNs in the co-added g or r band ZTF images, whose magnitude and redshift distributions are indicated by the orange color in panels (a)--(c) of Figure \ref{fig:AGNcat_dis}. On average, the depth of the 143 ZTF-derived AGNs is about 1 magnitude brighter in both Bmag and Rmag than that of the MQC's 205 AGNs.

To ensure the quality and completeness of the new light curves, we apply a filtering process to the 143 AGNs. The selection criteria are as follows: (1) all sources must be detected in both g and r bands of each co-added ZTF image; (2) an average S/N greater than 4; and (3) exclusion of sources located near the edges of the images. After the filtering process, we obtain a final sample of 73 AGNs in the EGS field (as indicated by the purple color in panels (a)--(c) of Figure \ref{fig:AGNcat_dis}). The subsequent analyses are based on these 73 AGNs.

We augment the physical parameters of our 73 AGNs by cross-matching them with the AGN Physical Parameter Catalog \citep{Szymon-2017ApJS} using a 1-arcsec matching radius. This catalog, derived from the SDSS Quasar Data Release 12 (DR12Q) \citep{SDSSDR12-2017A&A}, provides key parameters including black hole masses, bolometric luminosities, and Eddington ratios. Among our sample, 17 AGNs are matched and supplemented with these physical parameters. Additionally, we retrieve the ACSID information for all 73 AGNs from the EGS Photometric Catalog \citep{Davis-2007ApJ, Lotz-2008ApJ}. This catalog is based on HST/ACS photometric data obtained through the All-wavelength Extended Groth Strip International Survey (AEGIS), covering observations in both the F606W ($V$-band) and F814W ($I$-band) filters.

To characterize the variability behavior of the 73 AGNs, we calculate their variability amplitudes using the methods described in \citet{Sesar-2007AJ}, \citet{Ai-2010ApJL}, and \citet{Zuo-2012ApJ}:

\begin{equation}{} 
\Sigma = \sqrt{\frac{1}{n-1}\sum_{i=1}^{n}(m_{i} - \left \langle m \right \rangle )^{2} } ,
\label{equ-2}
\end{equation}

\begin{equation}{} 
\xi^{2} = \frac{1}{n} \sum_{i=1}^{n} (\Delta m_{i} )^{2} ,
\label{equ-3}
\end{equation}

\begin{equation}{} 
V = \begin{cases}
  (\Sigma ^{2} - \xi ^{2} )^{1/2} , & \text{ if } \Sigma > \xi  , \\
  0, & \text{ otherwise, } 
\end{cases}
\label{equ-4}
\end{equation} 
where n represent the count of exposures for a given source in each band, $m_{i}$ denotes the magnitude of the $i$-th observation, $\left \langle m \right \rangle $ indicates the average magnitude, and $\xi$ is calculated based on the photometric uncertainty $\Delta m_{i}$. This method mitigates the impact of photometric errors, providing a more accurate representation of AGN intrinsic variability.

We present a AGN catalog of 73 AGNs in the EGS field, consisting of 32 quasars and 41 Seyfert galaxies (see Table \ref{tab:2} for full details). As shown in panels (d)--(f) of Figure \ref{fig:AGNcat_dis}, the AGNs span a redshift range of 0-3.5 and exhibit variability amplitudes of 0.0-0.3 mag. For the subset of 17 AGNs (marked with a star symbol), we present the relationship between redshift and absolute magnitude, as well as black hole mass and bolometric luminosity. The complete catalog is available via the Appendix link.

\begin{table*}
\begin{center}
\renewcommand{\arraystretch}{1.2}
\setlength{\tabcolsep}{1.2pt}
\caption{The AGN catalog information of 73 AGNs in the EGS field.}
\label{tab:2}
\begin{tabular*}{\linewidth}{@{\extracolsep{\fill}}cccc@{}}
\hline
Column Name          & Format         & Unit                       & Description                                                                              \\ \hline
INDEX                & int64          &                            & Database ID for each AGN in master catalog                                               \\
ACSID                & int64          &                            & Database ID for each AGN in HST/ACS (from \cite{Davis-2007ApJ})              \\
RA                   & float64        & deg                        & Right ascension of the target                                                            \\
DEC                  & float64        & deg                        & Declination of the target                                                                \\
TYPE                 & str            &                            & Types of AGN (from \cite{Flesch-2023OJAp} )                             \\
Z                    & float64        &                            & Redshift (from \cite{Flesch-2023OJAp})    \\
log\_MBH             & float64        & $\log_{10}{(M_{\odot})}$   & $\log_{10}$ of the black hole mass  (from \cite{Szymon-2017ApJS} )      \\
log\_LBOL            & float64        & $\log_{10}{(erg s^{-1})} $ & $\log_{10}$ of the bolometric luminosity (from \cite{Szymon-2017ApJS} ) \\
log\_nEdd            & float64        &                            & $\log_{10}$ of the eddington ratio   (from \cite{Szymon-2017ApJS} )     \\
MAG\_g               & float64        & mag                        & Average magnitude in $g$-band                                                              \\
MAG\_ERR\_g          & float64        & mag                        & Average magnitude error in $g$-band                                                        \\
MAG\_g\_flag         & float64        & mag                        & Photometry type in $g$-band                                                            \\
MAG\_r               & float64        & mag                        & Average magnitude in $r$-band                                                              \\
MAG\_ERR\_r          & float64        & mag                        & Average magnitude error in $r$-band                                                        \\
MAG\_r\_flag         & float64        & mag                        & Photometry type in $r$-band                                                            \\
LC\_g                & array(float64) & mag                        & Light curve in $g$-band                                                                \\
LC\_ERR\_g           & array(float64) & mag                        & Light curve error in $g$-band                                                        \\
LC\_r                & array(float64) & mag                        & Light curve in $r$-band                                                                \\
LC\_ERR\_r           & array(float64) & mag                        & Light curve error in $r$-band                                                          \\
OBSJD                & array(float64) & day                        & JD of the observation                                                                    \\
VAR\_AMP\_g          & float64        & mag                        & Variability amplitude in $g$-band                                                      \\
VAR\_AMP\_r          & float64        & mag                        & Variability amplitude in $r$-band                                                      \\ \hline
\end{tabular*}
\end{center}
\end{table*}

\section{ANALYSES AND RESULTS} 
\label{3}
\subsection{Analysis of Optimized Light Curves}
\label{3.1}

\begin{figure*}[!htbp]
\centering
\includegraphics[width=1.9 \columnwidth]{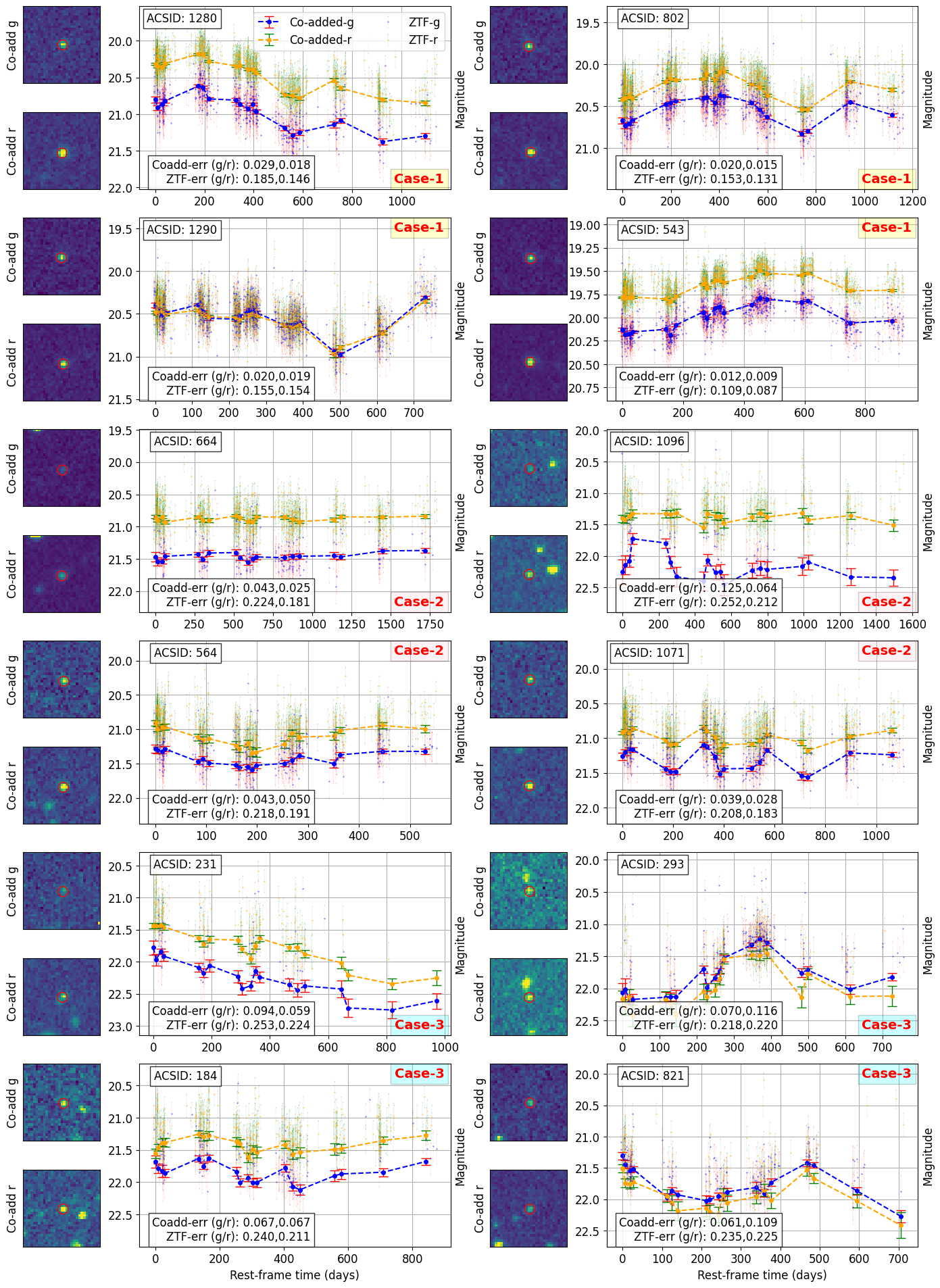}
\caption{Comparison of light curves. The thick orange and blue dots represent the $g$-band and $r$-band light curves extracted in this study, and the error bars are shown in red and green, respectively. The light dots correspond to the traditional ZTF light curves. Coadd-err ($g$/$r$) and ZTF-err ($g$/$r$) represent the average photometric uncertainties for the co-added and traditional ZTF light curves, respectively. The cutout of the co-added images is from Bin 1, with the red circle indicating the corresponding source. Day 0 in the light curve corresponds to the observation time of Bin 1, with a Julian Date of 2458213.36.}
\label{fig:lightcurve}
\end{figure*}

To evaluate the reliability of our newly co-added light curves, we compare them with traditional ZTF light curves. The traditional ZTF light curves are obtained from the NASA/IPAC Infrared Science Archive (IRSA; ZTF Team \citep{ajade68abib81}). We present the new light curves for several AGNs in Figure \ref{fig:lightcurve}. The $g$-band and $r$-band data are shown as thick orange and blue dots, respectively, with error bars in red and green. The traditional ZTF light curves are plotted as light dots for comparison. By comparing the light curves of four AGNs (ACSIDs 1280, 802, 1290, and 543, all exhibiting significant variability) in the top two panels of Figure \ref{fig:lightcurve}, we can see that the long-term variability trends of the new light curves align well with those of ZTF. Moreover, the new light curves exhibit higher precision. For example, for the AGN with ACSID 1280, the average uncertainties in the $g$-band and $r$-band from ZTF are 0.185 mag and 0.146 mag, respectively, while our new light curves reduce these to 0.029 mag and 0.018 mag. Similar improvements are observed for the other three AGNs. 
There are 8 AGNs in total exhibiting significant variability ($\Delta g >$ 0.1 mag and $\Delta r >$ 0.1 mag), with their light curve visualizations presented in ``Case-1'' of the Appendix.

As shown in the middle two panels of Figure \ref{fig:lightcurve}, the traditional ZTF light curves for four AGNs (ACSIDs 664, 1096, 564, and 1071) exhibit barely discernible variability due to large uncertainties. In contrast, the new light curves clearly reveal the variability characteristics of these sources. For instance, ACSID 664 displays variability in both bands during the rest-frame time interval of 500 to 750 days, with an amplitude significant at the 3$\sigma$ level. ACSID 564 shows a pronounced trend of decreasing then increasing brightness over the rest-frame time interval of 0 to 300 days. ACSIDs 1096 and 1071 exhibit significant variability throughout the entire rest-frame time, but these features are scarcely evident in the traditional ZTF light curves. The new light curves effectively capture these details, enhancing the detection of variability behaviors. Among the 73 AGNs, 29 show the above features, and their light curve visualizations are presented in ``Case-2'' of the Appendix.

Sources near or below the single-epoch detection limit typically lack usable light curves in the traditional ZTF data. Through our optimization, we reconstruct their previously undetected variability signatures. As presented in the bottom two panels of Figure \ref{fig:lightcurve}, four AGNs (ACSIDs 231, 293, 184, and 821) exhibit brightness fainter than 21 mag in both the $g$ and $r$ bands. The traditional ZTF light curves of ACSID 231 show sparse sampling in both bands during the rest-frame period of 600-1000 days. Our method can recover the missing photometric data while maintaining high precision. Using the new light curves, we can clearly observe that the brightness of ACSID 231 consistently declines over the rest-frame period of 600 to 1000 days. Similarly, our method supplements the $g$-band light curve data for ACSID 184. For ACSIDs 293 and 821, our method not only recovers the missing $r$-band light curve data but also clearly captures their long-term variability trends. The remaining 36 AGNs, about half of the AGN sample, exhibit signals near or below the detection limit of single-epoch images, with their light curve visualizations presented in ``Case-3'' of the Appendix.

In summary, only 11\% of the sources (8 AGNs, the "Case-1" sample) in this AGN sample show significant variability, which can be observed on both the traditional ZTF light curves and our newly co-added light curves. The new light curves of 40\% of the sources (29 AGNs, the "Case-2" sample) show a significant improvement in accuracy compared to the traditional ZTF light curves. The remaining 49\% of the sources (36 AGNs, the "Case-3" sample) have magnitudes close to or below the ZTF single exposure limit, making it difficult to detect their variability from traditional ZTF light curves. 

\subsection{Color Variations}
\label{3.2}

To investigate the long-term color variations of the 73 AGNs using our co-added light curves, we perform a linear regression analysis between the color index ($g$ - $r$) and the $g$-band magnitude, calculating the correlation coefficient $r$ and significance level $P$. Based on the correlation between color indices and magnitudes, the color variations of AGNs can be categorized into three types, i.e., colors become bluer as the source brightens (bluer-when-brighter, BWB), colors become redder as the source brightens (redder-when-brighter, RWB), or no clear correlation.

\begin{figure}
\center
\includegraphics[width=1 \columnwidth]{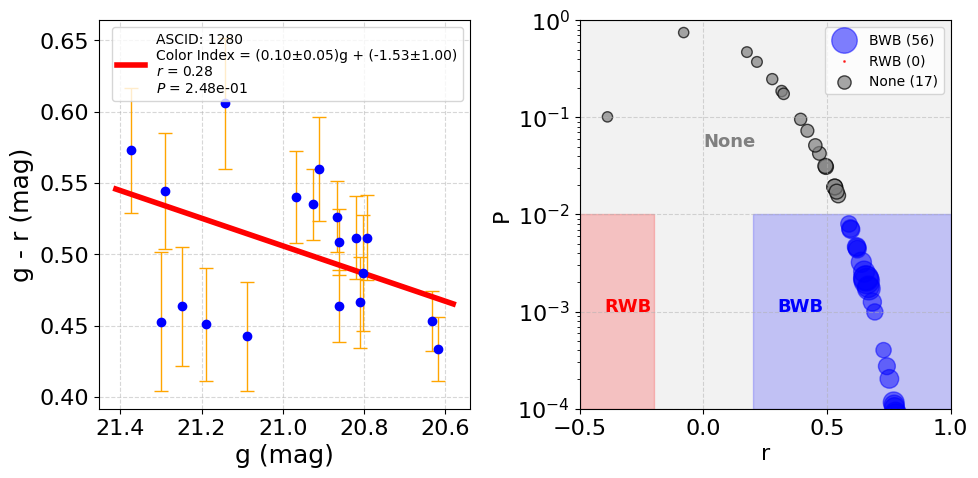}
\caption{Left: The relationship between the color index ($g$ - $r$) and $g$-band magnitude for AGN ACSID 1280. Right: The distribution of the correlation coefficient $r$ and significance level $P$ for 73 AGNs. The shaded regions are colored blue (indicating a BWB trend), red (indicating a RWB trend), and gray (indicating no significant BWB or RWB trend, labeled as None).}
\label{fig:color_behavior}
\end{figure}

Following the criteria of \citet{Zhang-2015RAA} and \citet{Li-2018RAA}, a positive correlation with $r > 0.2$ and $P < 0.01$ indicates a BWB trend, while a negative correlation with $r < -0.2$ and $P < 0.01$ indicates an RWB trend. All other cases imply no significant correlation between the color index and source brightness. As shown in Figure \ref{fig:color_behavior}, the left panel displays the color index ($g$ - $r$) versus the $g$-band magnitude for one AGN (ACSID 1280) over long-term timescales, while the right panel shows the distribution of the correlation coefficient $r$ and significance level $P$ for all AGNs. Among the 73 AGNs studied, 56 exhibit a BWB trend, while the remaining 17 show no significant correlation between color and brightness.

\section{Discussions} 
\label{4}

\subsection{Comparison with Rebinning Method for Faint AGNs}
\label{4.1}

To reduce uncertainties in light curve data, a common approach is to perform rebinning on photometric values in adjacent time windows \citep{Smith-2018ApJ}. Although this method is effective for light curves with high S/N, it fails to recover the true variability of faint AGNs whose brightness is near or below the detection limit in single observations. As shown in Figure \ref{fig:lightcurve}, the brightness of ACSID 231 is fainter than 22 mag in $g$-band and fainter than 21.5 mag in $r$-band. In the traditional ZTF light curves, the photometric data for this source are sparse, with large errors (average photometric errors of 0.253 mag in $g$-band and 0.224 mag in $r$-band), and some time periods lack any photometric data. Directly performing rebinning on the ZTF original photometric data for this source cannot recover its true variability signal. In contrast, our method stacks multi-epoch images to extend the effective exposure time and enhance image depth \citep{White-2007ApJ,Zibetti-2007ApJ,Zackay-2017ApJ-1,Zackay-2017ApJ-2}, which allow us to extract variability features of faint AGNs that are undetectable in single frames.

\subsection{Temporal Resolution} 
\label{4.2}

The temporal resolution of our light curves is reduced due to the stacking of at least 40 individual exposure images, spanning observational data from several days to months. While image stacking improves the S/N and enables deeper detection of faint sources, it reduces the ability to resolve short-term variability of AGNs. However, when shallower detection limits are acceptable, the temporal resolution can be enhanced by decreasing the number of co-added exposures.

We employ a uniform binning strategy for all sources within the same CCD quadrant in Section \ref{2.2.1}. Future work may implement optimized binning schemes, particularly for each AGNs with known variability properties (e.g., characteristic timescales from preliminary light-curve analyses or archival data), to better resolve their variability features.

\subsection{Color Behavior on Long-term Timescales} 
\label{4.3}

Among the 73 AGNs, the majority exhibit a BWB trend on long-term timescales. This result is consistent with previous studies reporting BWB dominance in AGN variability across both short- and long-term timescales \citep{Schmidt-2012ApJ,Zuo-2012ApJ,Sun-2014ApJ,Guo-2016ApJ,Hong-2017AJ,Li-2018RAA}. As suggested by \citet{Guo-2016ApJ}, the BWB trend is primarily driven by fluctuations in the inner accretion disk (e.g., localized thermal fluctuations \citep{Kelly-2009ApJ} or large-scale temperature variations \citep{Dexter-2011ApJ,Ruan-2014ApJ}), causing the spectrum to become bluer as brightness increases. In contrast, the RWB trend may arise from fluctuations in the outer disk with a delayed response in the inner region. Since most AGNs exhibit the BWB trend, fluctuations are more likely to originate in the inner disk and rarely in the outer disk.

\subsection{The Influence of Binning Schemes on Short-term and Long-term Variability} 
\label{4.4}

\begin{figure*}[!htbp]
\centering
\includegraphics[width=2.1 \columnwidth]{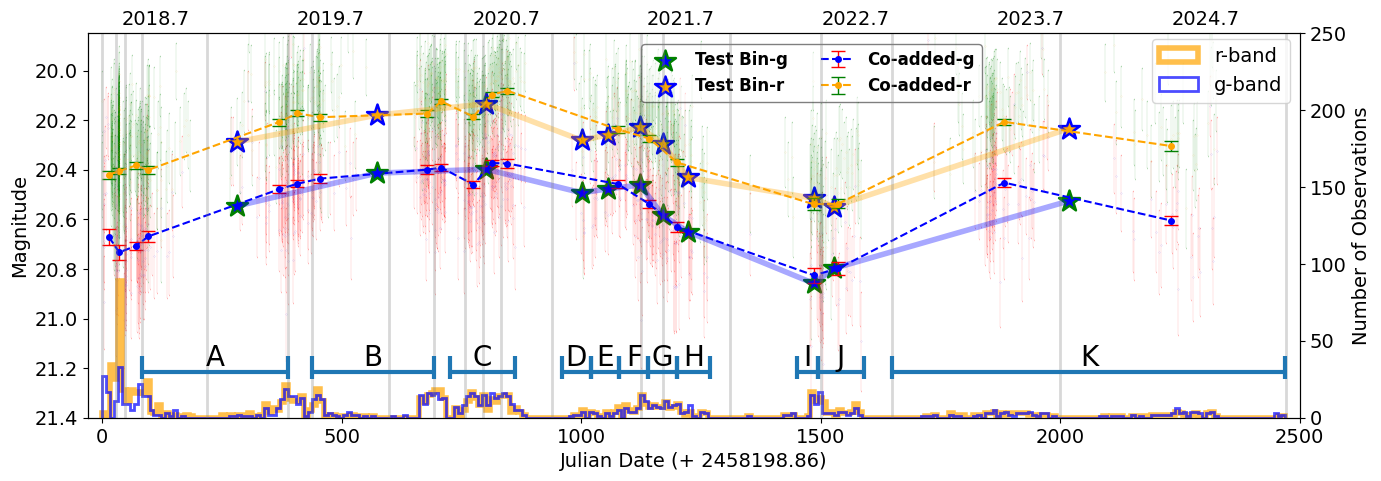}
\caption{The effects of bin placement and width on variability extraction for a sample light curve (ACSID 802), comparing the baseline scheme (gray vertical lines) against the test configurations (blue horizontal lines, A--K). The resulting light curves are shown as stars (test configurations), bold points (baseline co-added), and light points (traditional ZTF).}
\label{fig:newbin_lc}
\end{figure*}


We present the light curve of a typical source (ACSID 802) to investigate the influence of bin placement and width on the extraction of both short-term and long-term variability information. As shown in Figure \ref{fig:newbin_lc}, the gray vertical lines indicate the baseline bin intervals, whereas the blue horizontal lines (labeled A--K) represent a series of test configurations in which the bin start points and widths are varied. The resulting light curves from the test configurations (marked with stars) are compared against the baseline co-added light curve (bold points) and the traditional ZTF light curves (light points) to evaluate the sensitivity of variability measures to bin placement and width.

The choice of bin width and placement directly affects the preservation of short-term variability in our analysis of ACSID 802. As shown in Figure \ref{fig:newbin_lc}, wider bins (e.g., Bin C) smooth out fluctuations, while narrower bins (e.g., Bins D--H) preserve fine brightness changes. The binning process essentially averages the object's brightness variations over the period, yielding an composite brightness value (see Bins A, B, and K). Consequently, compared to short-term variability, the overall long-term trend remains robust across different binning configurations.

In summary, the choice of bin configuration is dictated by the source brightness and the scientific objective under a given observational cadence. For faint sources, wider bins are necessary to enhance the S/N, which preserves long-term trends but compromises temporal resolution. Conversely, for bright sources with high S/N, narrower bins can be employed to resolve finer variability, where the primary constraint is the observational cadence. As this study focuses on characterizing long-term variability in faint AGNs, our bin configuration is optimized for detection depth.

\section{SUMMARY} 
\label{5}

In this study, we present an image stacking method to optimize AGN long-term light curves, which is well suited to time-domain data from ZTF. By binning images according to observational cadence, we optimize the compromise between image depth and temporal resolution while ensuring sample completeness through consistent depth across all bins. Following \citet{Annis-2014ApJ}, we apply a weighting scheme based on seeing, sky brightness, and atmospheric transparency to enhance image co-addition. To validate the method, we construct a sample of 73 AGN light curves from ZTF observations using the deeper and more complete AGN catalogs in the EGS field as references. Compared to the traditional ZTF data, our method improves the quality of AGN long-term light curves, enabling better variability detection for faint AGNs. Next, we will extend this method to other ZTF fields to further construct high-precision optical light curves for faint AGNs and high-redshift AGNs.

Current and upcoming large-area time-domain surveys, such as the Wide Field Survey Telescope \citep[WFST;][]{WFST-2023SCPMA}, Euclid Space Telescope \citep{Euclid-2024arXiv}, Legacy Survey of Space and Time \citep[LSST;][]{LSST-2019ApJ}, and China Space Station Telescope \citep[CSST;][]{CSST-2011SSPMA}, will provide vast amounts of high-quality data. Building on this study's methodology, future research can better utilize survey data to improve measurement accuracy of AGN long-term variability and expand light curve samples for faint AGNs and high-redshift AGNs.



\begin{acknowledgments}
This work is supported by National Key R\&D Program of China No.2022YFF0503402. We also acknowledge the science research grants from the China Manned Space Project, especially, NO.  CMS-CSST-2025-A18. ZYZ acknowledges the support by the China-Chile Joint Research Fund (CCJRF No. 1906).

This work is based on observations obtained with the Samuel Oschin Telescope 48-inch and the 60-inch Telescope at the Palomar Observatory as part of the Zwicky Transient Facility project. ZTF is supported by the National Science Foundation under Grants No. AST1440341 and AST-2034437 and a collaboration including current partners Caltech, IPAC, the Weizmann Institute for Science, the Oskar Klein Center at Stockholm University, the University of Maryland, Deutsches Elektronen-Synchrotron and Humboldt University, the TANGO Consortium of Taiwan, the University of Wisconsin at Milwaukee, Trinity College Dublin, Lawrence Livermore National Laboratories, IN2P3, University of Warwick, Ruhr University Bochum, Northwestern University and former partners the University of Washington, Los Alamos National Laboratories, and Lawrence Berkeley National Laboratories.

\end{acknowledgments}

%







\facilities{ZTF}

\software{astropy \citep{2013A&A...558A..33A,2018AJ....156..123A,2022ApJ...935..167A}, SEP \citep{Barbary2016}, SExtractor \citep{Bertin-1996A&AS-sex}, PSFEx \citep{Bertin-2011ASPC-PSFEx}, SWarp \citep{Bertin-2010ascl}, TOPCAT \citep{2005ASPC..347...29T}, CARTA \citep{angus_comrie_2024_15172686}}


\begin{appendix}


\section*{Data Availability}
\label{sec:data}
The light curve data generated in this study are publicly available at \href{https://github.com/littlejiaqi/AGN\_catalog.git}{github repository}. The repository contains \href{https://github.com/littlejiaqi/AGN_catalog/blob/main/AGN_catalog.parquet}{the AGN catalog}, which includes data for 73 AGNs in the EGS field. The structure of the AGN catalog is detailed in Table~\ref{tab:2}. Additionally, we provide visualizations of three case of light curves, presented in the following HTML files: \href{https://github.com/littlejiaqi/AGN_catalog/blob/main/Case-1\_AGNs\_light\_curve.html}{Case-1} (containing 8 AGNs), \href{https://github.com/littlejiaqi/AGN_catalog/blob/main/Case-2\_AGNs\_light\_curve.html}{Case-2} (containing 29 AGNs), and \href{https://github.com/littlejiaqi/AGN_catalog/blob/main/Case-3\_AGNs\_light\_curve.html}{Case-3} (containing 36 AGNs).
\end{appendix}



\bibliography{sample7}{}
\bibliographystyle{aasjournal}


\end{CJK*}
\end{document}